# Evidence of electronic cloaking from chiral electron transport in bilayer graphene nanostructures

*Kyunghoon Lee[†, §], Seunghyun Lee[†, #], Yun Suk Eo[‡], Cagliyan Kurdak[‡], Zhaohui Zhong[†, *]*.

[†]Department of Electrical Engineering and Computer Science, University of Michigan, Ann Arbor, MI 48109, USA.

[‡]Department of Physics, University of Michigan, Ann Arbor, MI 48109, USA.

[§]Department of Electrical Engineering and Computer Science, University of California, Berkeley, CA 94720, USA.

[#]Department of Electronics and Radio Engineering, Kyung Hee University, Gyeonggi 446-701, South Korea.

**ABSTRACT**

The coupling of charge carrier motion and pseudospin via chirality for massless Dirac fermions in monolayer graphene has generated dramatic consequences such as the unusual quantum Hall effect and Klein tunneling. In bilayer graphene, charge carriers are massive Dirac fermions with a finite density of states at zero energy. Because of their non-relativistic nature, massive Dirac fermions can provide an even better testbed to clarify the importance of chirality in transport measurement than massless Dirac fermions in monolayer graphene. Here we report electronic cloaking effect as a manifestation of chirality by probing phase coherent transport in CVD bilayer graphene. Conductance oscillations with different periodicities were observed on extremely narrow bilayer graphene heterojunctions through electrostatic gating. With a Fourier analysis technique, we identify the origin of each individual interference pattern. Importantly, the electron waves on two sides of the potential barrier can be coupled through the evanescent waves inside the barrier, making the confined states underneath the barrier invisible to electrons. These findings provide a direct evidence for the electronic cloaking effect, and hold promise for the realization of pseudospintronics based on bilayer graphene.

Since the experimental observation of unconventional integer quantum Hall effects [1-3] in monolayer graphene, chirality has been considered as holding a key role in understanding unusual transport behaviors of massless Dirac fermions [4-8]. One prime example is the Klein tunneling, perfect transmission through the barrier regardless of its width and energy height [5,9-19]. Different from monolayer graphene, charge carriers in bilayer graphene (BLG) are massive Dirac fermions also with a chiral nature, but a finite density of states at zero energy [5,9,20,21]. The case of chiral tunneling in BLG bipolar junctions would be even more interesting [22-24], where complete decoupling between quasiparticle states of opposite polarity has been predicted theoretically at normal incidence due to chirality mismatch [9,23-25]. A striking consequence of chirality mismatch is the rendering of confined states via potential barrier from opposite pseudospin states invisible to electron transport – the so called electronic cloaking effect [23]. This electronic cloaking is different from optical cloaking in a sense that the probing waves directly tunnel through the potential barrier where cloaked states are contained, not by moving around cloaked objects as in optical cloaking effect [23,26,27]. However, experimental verification of the electronic cloaking effect is still challenging. To this end, we present the experimental evidence of electronic cloaking and Klein tunneling of massive Dirac fermions in BLG by probing the phase coherent transport behavior of dual gated BLG transistor.

The devices were fabricated on CVD bilayer graphene with channel lengths between 50 to 200 nm [28] (see Supplementary Information for detail). The schematic and a scanning electron microscopy image of representative devices are shown in Figs. 1a and 1b. In contrast to earlier implementation of dual gated devices, which induce Fabry-Perot interference only inside a potential barrier [13,29,30], the dual-gated BLG device described here allows a phase coherent transport regime over the full channel length, necessary condition to realize electronic cloaking phenomena. This approach allows us to prevent loss of phase information when charge carriers traverse the BLG channel.

Conductance oscillations arising from Fabry-Perot interference in these BLG devices can originate from several possible trajectories related to the chiral massive fermions. We first look at the monopolar regime (Fig, 1c). Graphene underneath the source and drain metal contacts can be doped by charge transfer from metals, and may have a different polarity from that of the channel region [31,32]. Fabry-Perot interference can appear with a resonance cavity length defined by the effective channel length of the device (Fig. 1c) [13,32-34].

In a more sophisticated way, the bipolar regime, where the polarity of graphene regions controlled by the back gate and the top gate are different, can offer an opportunity to unravel crucial role of chirality in transport through the potential barrier [13,14,16-18]. As shown in Fig.

1d, conductance oscillations can arise from electron (blue arrow) and hole (yellow arrow) round-trip resonances confined within the back gate controlled left and right graphene regions, and center region controlled by both top and back gates. More interestingly, chiral carriers could have additional possible route to complete a round trip across the potential barrier via quantum tunneling (green arrow), as if the confined states underneath the barrier are invisible to the carrier transport. At normal incidence, the coupling between the positive and negative energy states is completely suppressed due to pseudospin conservation in bilayer graphene, in which chirality is tied to the polarity of charge carriers [9,23]. The quasiparticle states on both side of the barrier, however, have the same pseudospin and can be coupled via evanescent waves. As the results, the potential barrier acts as a cloak for the underneath confined states with opposite pseudospin, rendering them invisible to the massive Dirac fermion transport across the barrier. The observation of resonance across the entire device in bipolar regime will thus provide a strong proof for the intriguing electronic cloaking effect in BLG.

We first characterize carrier transport in single back-gated BLG devices via two-terminal measurements at 6 K. Fig. 2a shows the two dimensional color plot of differential conductance as a function of back-gate voltage ($V_{BG}$) and bias voltage ($V$) for a device with 55 nm channel length. Smooth background is subtracted to enhance the periodic patterns. Chessboard patterns are clearly visible in the 2D plot, with the quasi-periodic patterns highlighted by the guide lines. Similar periodic features can also be observed in devices with 118 nm and 160 nm channel lengths (Figs. 2b and 2c). These results are in agreement with the Fabry-Perot like quantum interference of electron waves [13,33,34].

The observed quantum interference can be understood by examining the round trip resonance condition. Whenever phase change obtained by round trip of an electron reaches 2π, a constructive interference pattern appears for $\Delta k_F = \pi/L$. Combining this condition with parabolic band dispersion of BLG gives,

$$eV_C = \frac{\hbar^2 \pi C_{BG} \Delta V_{BG}}{2m^*} = \frac{\hbar^2}{m^*}\left[\left(\frac{\pi}{W}\right)^2 + \left(\frac{\pi}{L}\right)^2\right] \approx \frac{\hbar^2 \pi^2}{m^* L^2} \qquad (1)$$

where m* is the effective mass having $0.03 m_e$ [33], and $C_{BG}$ is the gate capacitance of 175 nF/cm$^2$ calculated from parallel plate capacitor model. We note that here we only consider the lowest energy mode of Fabry-Perot oscillation. The oscillation periods, in $eV_C$, scale inversely with the square of the channel length (L) for BLG. In comparison, for single layer graphene, due to its linear band dispersion, the $eV_C$ is expected to be inversely proportional to L. Moreover, gate voltage oscillation period ($\Delta V_{BG}$) is similar at low and high carrier density for BLG, whereas $\Delta V_{BG}$ changes significantly as gate voltage varies for SLG [33]. We further extract the bias voltage oscillation period ($V_C$) and gate voltage oscillation period ($\Delta V_{BG}$), respectively, for three devices: 21(±2) mV, 0.45(±0.02) V for the L = 55 nm device; 5.1(±0.3) mV and 0.12(±0.02) V for the L = 118 nm device; 3.1(±0.1) mV and 0.065(±0.02) V for the L = 160 nm device. As shown in Fig. 2d, the measured $eV_C$ scales inversely with $L^2$, and the small deviation can be attributed to the fact that the resonance cavity length is smaller than its physical length due to fringing field screening effect from metal electrodes [35]. We estimate an resonance cavity length of 50 nm for the device with 55 nm physical length. These results also indicate that the phase coherence length in our CVD BLG is larger than 160 nm.

To study the anomalous chiral electron tunneling behavior, we now focus on the transport measurements of the dual-gated BLG devices, starting with a 150 nm channel length device. The differential resistance is measured as a function of back and top gate voltages. Figure 3a shows the 2D color plot of the resistance, and the four quadrants correspond to the monopolar (n-n-n and p-p-p) and bipolar regimes (p-n-p and n-p-n). The slope of charge neutral line gives capacitive coupling ratio between top-gate and back-gate, $C_{TG}/C_{BG}$ ~1.7. A rich set of oscillatory features is observed in our differential resistance map; one interference pattern in the monopolar regime, and more than two fringes forming checkerboard like complex interference patterns in the bipolar regime.

In order to understand the origin of the resonances and gain a better insight into the transport mechanism, we employ a 2D Fourier analysis technique for extracting the interference patterns [36]. The main role of 2D FFT technique is to separate fringe patterns affected by different combinations of top and back gate voltages along the horizontal ($N_B$) and diagonal ($N_T$) orientations. By masking one fringe pattern and performing inverse FFT on the other fringe patterns, we can clarify the presence of two different fringe components along each orientation (Figs. 3b and 3c). The summation of these separated fringe patterns recovers the original 2D differential resistance map with more pronounced interference patterns (Fig. 3d).

The FFT filtered data sets enable us to better understand observed interference patterns. We first examine the resonances along the $N_T$ orientation as shown in Fig. 3b. In the monopolar region, sequences of periodic oscillations are clearly visible as indicated by purple color line. The FFT along $V_{BG}$ in the monopolar region shows a peak frequency at 16.39 (1/V), corresponding to oscillation period of $\Delta V_{BG} \sim 0.061$ V (see Fig. S2). Using equation (1), this value corresponds to a BLG resonator cavity length of ~ 138 nm, in agreement with the physical length between source and drain electrodes (150 nm). By calculating the mobility as $\mu = \frac{\sigma}{ne}$ and the mean free path as $l_e = \frac{h}{e^2}\frac{\sigma}{2\sqrt{\pi n}}$, we estimated carrier mobility, $\mu \sim 4000$ cm$^2$V$^{-1}$s$^{-1}$ with $l_e \sim 82$ nm at carrier density of $3 \times 10^{12}$ cm$^{-2}$. We note that this value represents the lower bound mean free path without removing the effect of contact resistance [37]

We now turn our attention to the bipolar region. Oscillations along $N_T$ direction with different periodicities are clearly present, as highlighted by the green and dotted yellow lines in Fig. 3b. The FFT analysis in the bipolar region (Fig. 3e) yields two primary oscillation periods of 0.270 V, and 0.151 ~ 0.189 V, which correspond to resonant cavity length of 65 nm, and 78 ~ 87 nm, respectively. The cavity length of 65 nm is in reasonable agreement with the fabricated top-gate width of 30 nm, taking into account of smooth potential profile [13,14,16,38] due to the estimated width of p-n junction, ~18 nm (see Fig. S3 ). This result is consistent with the existence of localized state confined underneath the top-gate by potential barriers (yellow arrows in Fig. 3b inset).

The second oscillation, corresponding to cavity sizes of 78 ~ 87 nm, is more intriguing. The resonant condition for these cavities immediately invites one to consider electrons bouncing back and forth within graphene sections between top-gate and metal electrodes; states confined in either left or right BLG lead regions. However, we rule out this possibility because 1) density of carriers localized in BLG leads can only be tuned by back-gate voltage, not by top-gate voltage [39], and 2) the length of the resonant cavity (78 ~ 87 nm) is also larger than the resonant cavity length of the left and right BLG leads (~ 60 nm) by considering width of p-n junction (~ 18 nm). The other plausible explanation is that plane waves, either on left and right side of potential barriers, are coupled through the evanescent waves inside the barrier, as if there are no available states under the barrier (green arrows in Fig. 3b inset). Note that energy of normally incident quasiparticle waves is slightly lower than the barrier height, but high enough to tunnel through the barrier, for the realization of cloaking effect. Otherwise, this condition is not satisfied since the tunneling probability decreases significantly as energy difference increases between quasiparticle waves and the barrier height. The observation of resonance mode with trajectory bouncing through the full device is highly unusual, thus demonstrating that the necessary condition is fulfilled. Indeed, the value of full channel length (~150 nm) approximately equals the sum of the resonant cavity length defined by the top gate (~ 65 nm) and the quasiparticle propagation distance (78 ~ 87 nm). The results thus provide the experimental evidence for the electronic cloaking effect due to the pseudospin mismatch of opposite polarity. The discrepancy of propagation distance, ~ 9 nm, is attributed to the variation of potential profile as back-gate voltage sweeps. The splitting of resonant peak in FFT analysis can be understood in the context of this potential variation. This interpretation is carefully investigated by identification of corresponding oscillations in original 2D color map, and is also confirmed with density profiles using finite-element simulation software (Fig. S3). Rather small variation of potential profile seems to be due to electric field screening of back-gate by top-gate, source, and drain metal electrodes in our BLG nanostructures [31,40,41].

We also study the oscillations along $N_B$ direction (Fig. 3c), where periodic resonance is clearly visible in the bipolar region. The FFT spectrum (Fig. 3f) taken from the data along the dashed line in Fig. 3c show an oscillation period of 0.391 V, corresponding to a resonant cavity length of 54 nm. The resonant cavity length is consistent with the physical length of left BLG lead (~ 80 nm,

Fig. 3f inset) by taking account of width of p-n junction (~ 18nm) due to smooth potential profile. The result indicates that a significant fraction of near normally incident waves with energy much smaller than the barrier height are reflected at the interface of the p-n junction. Given the confined states underneath the top-gate, we attribute this feature to the chirality mismatch leading to the suppression of electron-hole coupling, consequence of anti-Klein tunneling effect [9]. Regarding resonant cavity of right BLG lead, expected quantum interference was not observed neither on FFT or on 2D resistance map due to the fact that oscillation period seems to be beyond our detection limit, ~2V, in terms of back-gate voltage (Figs. 3f inset and 3g inset). The slight asymmetry of right and left BLG lead is due to the shift of top-gate by 15 ~ 20 nm, confirmed from the SEM image (Fig. 3f inset).

We now note that cloaking resonance cavity should be equal to the sum of resonant cavity length of left and right BLG leads. Even though measured oscillation peak was not found for right BLG lead cavity, we still could estimate resonant cavity size (~22 nm) by subtracting width of p-n junction from physical length of right BLG lead (~ 40 nm) determined from SEM image (Fig. 3f inset). Given this, and in light of analysis on cloaking cavity, the sum of estimated resonant cavity length of left and right BLG leads, 76 nm, is still quantitatively in good agreement with cloaking cavity size within error range ($\approx$ 10 nm) owing to potential profile variation as back gate voltage changes. Such remarkable agreement provides further evidence that the observed cloaking resonance, full round trip across the BLG channel by tunneling the barrier, results from the coupling between quasiparticles with same pseudospin orientation.

Similar quantum interference phenomena were observed from other devices with 120 nm and 100 nm channel length, respectively (see Supplementary Information). Importantly, in the bipolar region, they also show oscillation features corresponding to 1) resonance cavity defined by the top-gate, and 2) a "cloaking cavity" extended throughout the whole device with states underneath the top-gate been invisible. For all samples, we carefully record FFT results along vertical lines at 10 different top-gate voltages within highlighted dashed square regions both at bipolar and monopolar regimes (Table 1). As such, we also identify low frequency peaks as noise signal (Figs. 3e, 3f, and 3g) because those were not repetitively present at different FFT results, and the corresponding periods were not identified in the original 2D resistance map. The results from all three devices paint a consistent picture of different types of quantum interference due to the chiral nature of massive Dirac fermion, as depicted in Fig. 1d.

In conclusion, we present experimental evidence of electronic cloaking effect by probing phase-coherent transport behavior in our CVD BLG devices. Further evidence can come from magnetic field dependent studies [13,30], and by fabricating device on *h*-BN substrate [19,42] to achieve much longer phase coherence length. The results suggest the utilization of chiral dependent transport properties to encode information using pseudospins of massive fermions, and may pave the way for future applications of pseudospintronics with bilayer graphene.

FIGURES

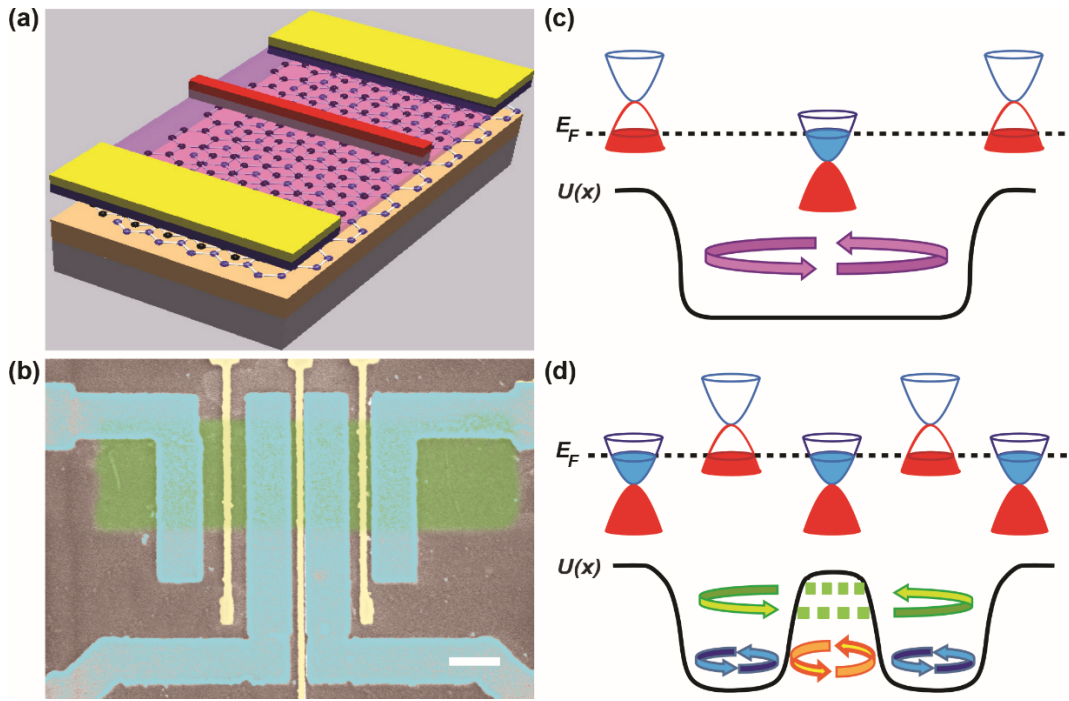

**FIG. 1** (a) Schematic diagram of a dual-gated BLG device. (b) False-color scanning electron microscopy image of three BLG devices having different channel length (200, 50, and 100 nm from left to right). Source/drain electrodes, BLG, and top-gate electrodes are represented by blue, green, and yellow color, respectively. The scale bar is 200 nm. (c) Schematic showing Fabry-Perot resonance in monopolar regime due to reflection at the source and drain contact. (d) Schematic illustration of electronic cloaking resonance in BLG arising from decoupling of orthogonal pseudospins in bipolar regime (green color). Blue and yellow colored trajectories show resonance of confined states by potential barrier.

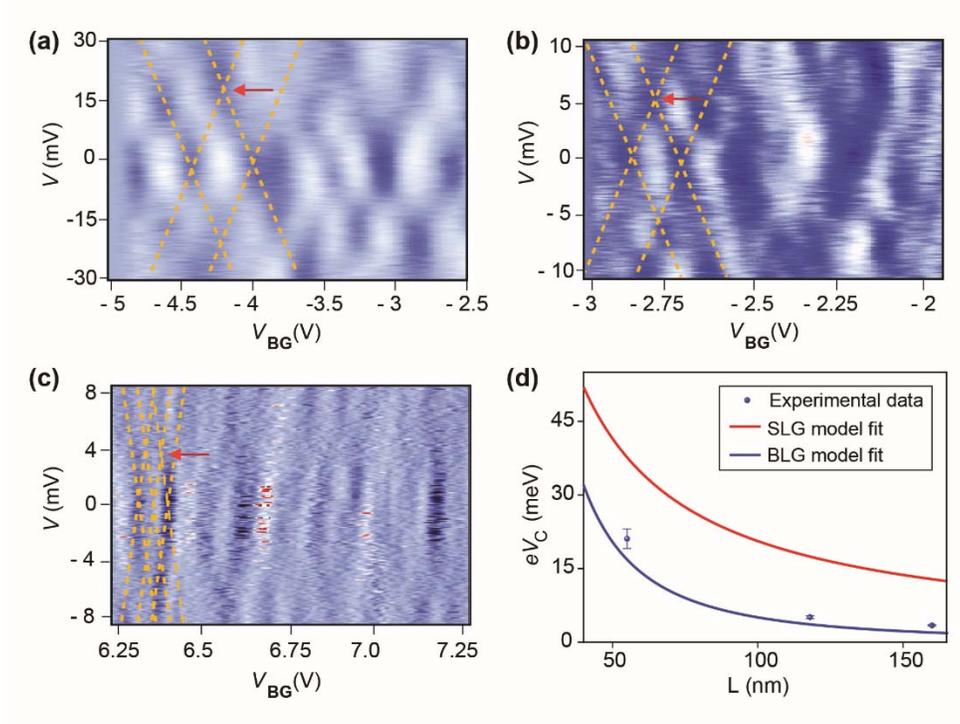

**FIG. 2** Fabry-Perot resonance in back-gated BLG devices. (a), (b), and (c), Two-dimensional color plot of differential conductance versus V and $V_{BG}$ for BLG device with channel length of (a) 55 nm, (b) 118 nm, and (c) 160 nm. The channel width is 500 nm for all three devices, and all data are taken at 6K. For each image, a smooth background was subtracted to highlight the Fabry-Perot oscillation patterns. Quasi-periodic crisscrossing dark (bright) lines correspond to conductance dips (peaks). The dotted yellow lines are guides to the eye, and the red arrows indicate the bias voltage oscillation period ($V_C$). (d) $eV_C$ measured from three devices plotted against device physical channel lengths. The results are also compared with theoretical length dependent resonance periods for SLG with linear dispersion (red curve) and BLG with parabolic dispersion (blue curve).

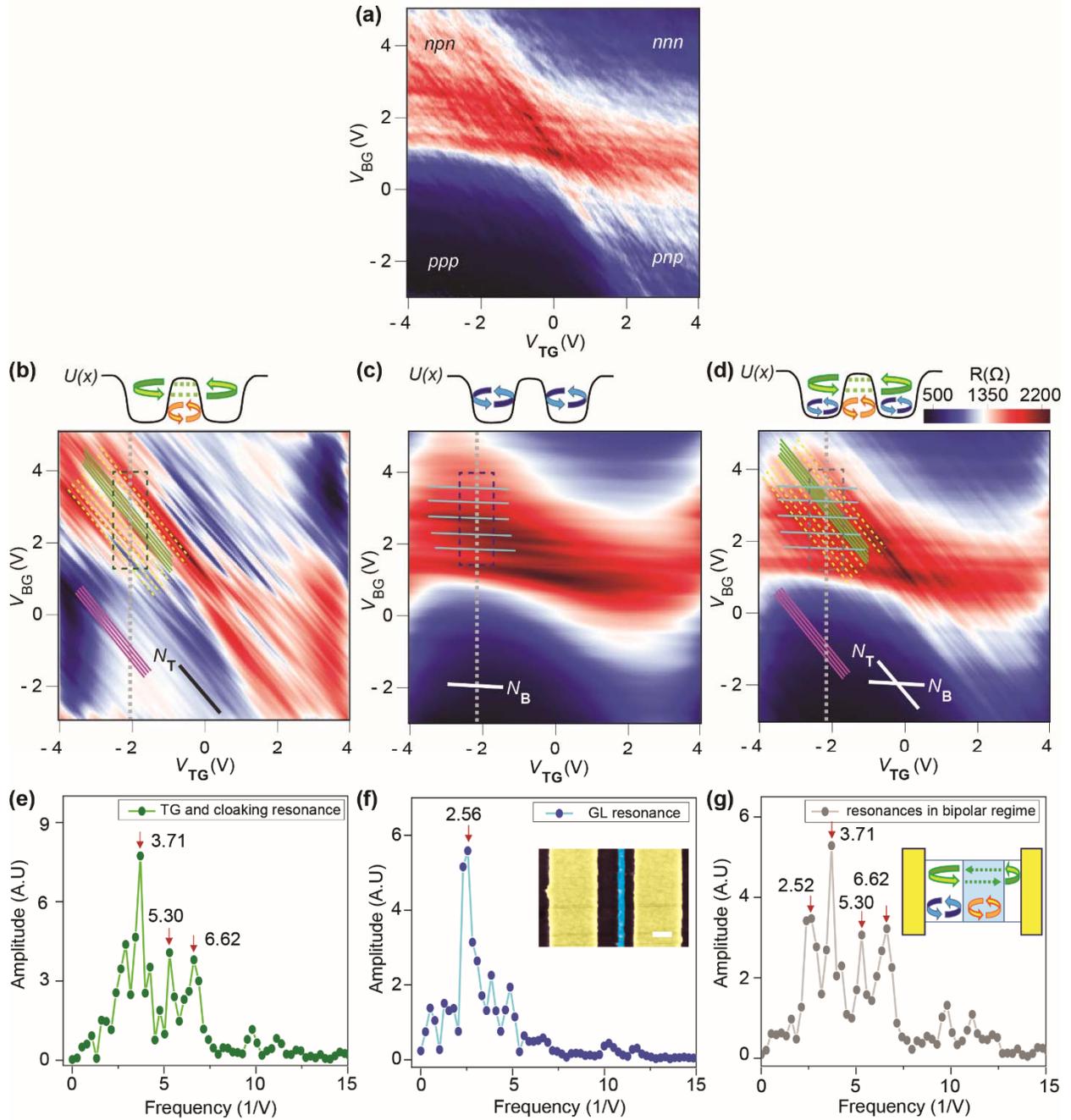

**FIG. 3** Electronic cloaking effect in dual-gated BLG device. (a) Two-dimensional resistance map as a function of $V_{TG}$ and $V_{BG}$ from a 150 nm BLG device. (b) and (c), To extract oscillation components controlled by both $V_{TG}$ and $V_{BG}$, and controlled only by $V_{BG}$ separately, inverse Fourier transform with masking technique was performed on 2D resistance map in (a). The two fringing

patterns along (b) $N_T$ (black solid line) and (c) $N_B$ directions (white solid line) were obtained respectively. (d) Two-dimensional resistance map is recovered with clearer oscillation patterns by adding two fringing patterns, (b) and (c). Top insets in (b), (c), (d) show representative trajectories for each resonance condition. Observed Fabry-Perot conductance oscillations are represented by yellow, green, and blue lines, respectively in (b), (c), (d), corresponding to massive Dirac fermions trajectories with the same colors shown in the insets. (e), (f), (g), The Fourier transform spectra for (b), (c), and (d), respectively. The FFTs are performed along the grey dashed line ($V_{TG} = -2.1V$) in the highlighted dashed square regions. Note that weak higher harmonics are also observed, corresponding to multiple rotating trajectories within the defined cavities. The inset in (f) shows false-color SEM image of a device with top-gate (blue), source-drain electrodes (yellow), and BLG (dark purple). The scale bar is 100nm. The schematic view of resonant cavity lengths with trajectories is drawn for corresponding FFT results (inset in (g)).

TABLES.

| Channel Length | TG | | GL | | Cloaking | | | |
|---|---|---|---|---|---|---|---|---|
| Physical [nm] | Resonant cavity | | Physical [nm] | Resonant cavity | | Physical [nm] | Resonant cavity | | Resonant cavity | |
| | [nm] | $\Delta V_{BG}$(mV) | | [nm] | $\Delta V_{BG}$(mV) | | [nm] | $\Delta V_{BG}$(mV) | [nm] | $\Delta V_{BG}$(mV) |
| 150 | 136±3.2 | 60±3 | 30 | 64.4±2.5 | 280±20 | 80 | 54±2.2 | 390±30 | 82.3±3.5 | 170±10 |
| 120 | 109±8 | 98±15 | 30 | 50.1±5 | 470±90 | 60 | 50.2±4.4 | 460±70 | 74.6±3.2 | 210±20 |
| 100 | 94±7 | 130±20 | 30 | 44±3.3 | 640±110 | 50 | 37.7±3.7 | 830±160 | 58.3±3.7 | 340±40 |

**Table 1.** Physical lengths and resonance cavity lengths estimated from oscillation periods ($\Delta V_{BG}$) in the studied BLG devices.

We thank J. Song and L. Levitov for their helpful discussions. This work was supported by the National Science Foundation Scalable Nanomanufacturing Program (DMR-1120187), Z. Z.

acknowledge the support from NSF CAREER Award (ECCS-1254468). K.L. acknowledges support from Kwanjeong Educational Foundation. Devices were fabricated in the Lurie Nanofabrication Facility at the University of Michigan, a member of the NSF National Nanotechnology Infrastructure Network.